\documentclass{JINST}
\let\ifpdf\relax

 \usepackage{url}
\usepackage{epstopdf}
\usepackage{verbatim}
\usepackage{cite}

\usepackage{listings}
\usepackage{courier}
\RequirePackage{lineno}

\usepackage{caption}

\title{Accelerated Event-by-Event Neutrino Oscillation Reweighting with Matter Effects on a GPU}

\author{R G. Calland$^a$, A. C. Kaboth$^b$, D. Payne$^a$

\\
\llap{$^a$}University of Liverpool,
Department of Physics, 
  Oliver Lodge Bld,
Oxford Street,
Liverpool,
L69 7ZE, 
UK \\
\llap{$^b$}Department of Physics, Imperial College London, London, SW7 2AZ, UK

E-mail: \email{rcalland@hep.ph.liv.ac.uk}}

\abstract{Oscillation probability calculations are becoming increasingly CPU intensive in modern neutrino oscillation analyses. The independency of reweighting individual events in a Monte Carlo sample lends itself to parallel implementation on a \emph{graphics processing unit}. The library \texttt{Prob3++} was ported to the GPU using the CUDA C API, allowing for large scale parallelized calculations of neutrino oscillation probabilities through matter of constant density, decreasing the execution time by 2 orders of magnitude when compared to performance on a single CPU.}


\begin{document}


\section{Introduction}\label{sec:xxx}
Current and future long-baseline experiments are designed to observe an appearance or disappearance of neutrino events by studying a neutrino beam at various distances from the beam origin. This difference can be quantified by comparing the observed spectra to the non-oscillation case. To do this, a probability distribution function (PDF) must be constructed empirically from detector Monte Carlo (MC) and reweighted according to the neutrino oscillation model chosen and any corresponding systematic uncertainties.

\subsection{Neutrino Oscillation Probability}\label{sec:yyy}
\begin{figure}
\centering
\includegraphics[scale=0.7]{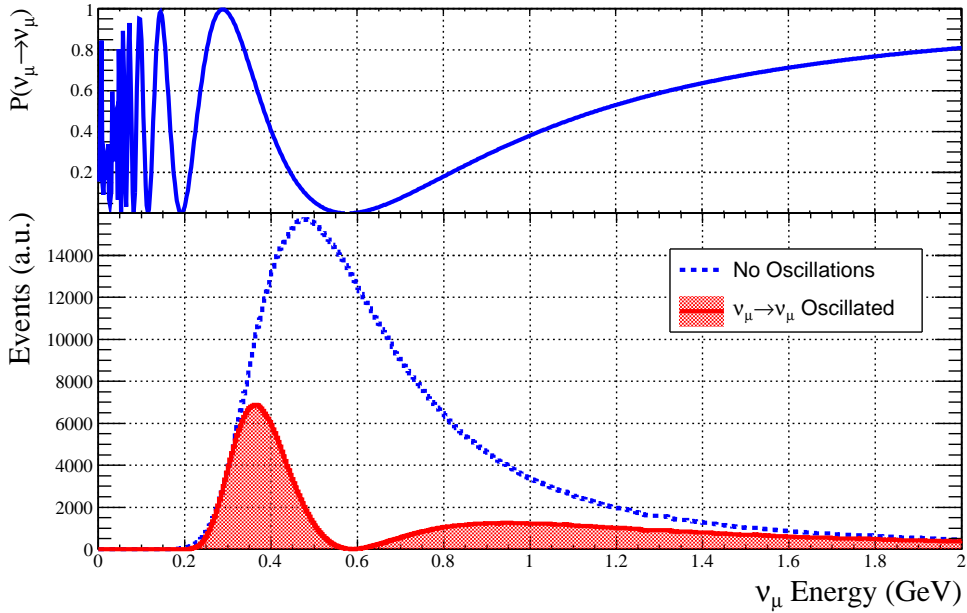}
\caption{\textbf{Top}: $\nu_\mu \rightarrow \nu_\mu$ neutrino survival probability calculated with matter effects for a propagation distance of 295 km through a constant matter density of 2.6 $g/cm^3$ . \textbf{Bottom}: A mock $\nu_{\mu}$ neutrino beam spectra under the influence of this oscillation probability, compared to the no oscillation case. The trough of the oscillation probability function can been seen to line up with the trough of the oscillated spectra at 0.6 GeV. Oscillations were calculated using parameter values listed in table~\protect\ref{tab:osc_table}  with normal hierarchy.}
\label{fig:oscprob}

\end{figure}

In the standard 3 neutrino formulation, neutrinos propagate as a superposition of three mass eigenstates  $m_{1,2,3}$. A neutrino interaction is governed by its flavour, and can be inferred indirectly via observation of the outgoing lepton from a neutrino interaction vertex. The probability that a neutrino of flavour $\nu_{\alpha}$ and energy $E$ (GeV) will be be observed with a flavour $\nu_{\beta}$ after propagation of distance $L$ (km) through vacuum can be determined from its mass states $m_{i}$ and the unitary PMNS transition matrix $U_{flavour,mass}:$

\begin{equation}
P(\nu_{\alpha} \rightarrow \nu_{\beta}) =  \left| \sum\limits_{i=1}^3 U_{\alpha i} \exp\left(-\frac{1}{2} i m^2_i \frac{L}{E}\right) \right|^2
\label{prob}
\end{equation}
This equation is illustrated for the $\nu_{\mu} \rightarrow \nu_{\mu}$ survival probability in the top plot of figure~\ref{fig:oscprob}.

The propagation of neutrinos through matter induces non-negligible effects on $\nu_e$ and $\bar{\nu}_e$ due to forward scattering on electrons in matter. These so-called matter effects add computational complexity but can be calculated as prescribed in~\cite{bib1}.

\begin{table}
\caption{Assumed oscillation parameters for all studies presented.}
\label{tab:osc_table}
\begin{center}
\begin{tabular}{|c|c|}
\hline 
Parameter & Value \\ 
\hline 
$\sin^{2}(\theta_{12})$ & 0.311 \\ 
\hline 
$\sin^{2}(\theta_{23})$ & 0.5 \\ 
\hline 
$\sin^{2}(\theta_{13})$ & 0.0251 \\ 
\hline 
$\Delta m^{2}_{32}$ ($eV^2$) & $2.4 \times 10^{-3}$ \\ 
\hline 
$\Delta m^{2}_{12}$ ($eV^2$) & $7.6 \times 10^{-5}$ \\ 
\hline 
$\delta_{cp}$ & 0 \\ 
\hline 
Earth Density ($g$/$cm^{3}$) & 2.6 \\ 
\hline 
Baseline ($km$) & 295 \\ 
\hline 
\end{tabular} 
\end{center}
\end{table}


\subsubsection{Event-By-Event Reweighting}
Neutrino oscillation analyses are often performed by producing a large sample of simulated events in order to estimate the PDF, as many reconstruction effects may be complex. These simulated events are produced at a certain set of oscillation parameters and experimental parameters, all of which must be varied in order to find the optimal output parameters for analysis. Binned maximum likelihood analyses are an effective way to compare the data to the MC to optimize the parameters. Calculating the effect of the variation of oscillation and systematic parameters can be done in two ways for these binned MC PDFs. One option is to calculate the effect of the variation at the center of each bin and apply it to the whole bin; this has the advantage of being relatively quick, but the disadvantage of losing any shape information which resides inside the bin boundaries. The other option is to retain all of the simulated events and calculate the variations on an event-by-event basis; this has the advantage of retaining any shape information within the bin, but the disadvantage of requiring many more calculations. 

Both oscillation parameters and systematic uncertainty parameters are subject to this binning effect. An example of a systematic uncertainty that would be impacted by binning is a scale uncertainty for energy reconstruction, critical for oscillation analyses. Using a binned weighting method loses the information about the reconstructed energy of any given event, and so produce a different predicted number of events than simply scaling the true reconstructed energy of the constituent MC events. Further discussion of systematic uncertainties is beyond the scope of this note, but it comprises part of the motivation to find a computationally efficient way to treat the constituent MC events individually. 

The binning effect on oscillation parameters can be as large as a few percent. One can see this effect by placing an histogram bin with a typical width of 25 MeV from 0.6 GeV to 0.625 GeV (near to the oscillation maximum shown in figure~\ref{fig:oscprob}). Considering the case of integrating the true neutrino energy spectrum in this bin and multiplying by the oscillation probability at the bin center (0.6125 MeV), and comparing this with the result of integrating the product of the oscillation probability and the input neutrino spectrum one finds a difference of 2.6\%. This difference arises from the approximation that all neutrinos within the bin edges have the same true energy.

 This is a strong motivation to find a way to treat the constituent MC events according to their true properties. Since this method increases the number of oscillation weight calculations by several orders of magnitude, it is not practical to perform these calculations on a CPU, and so we describe the implementation of this calculation on a GPU.


\section{Implementation on a GPU}
A typical CPU consists of $\sim4$ cores with clock speeds in the range of 3-4 GHz and have the capacity to run multi-threaded applications. In contrast, a modern consumer GPU has 100-1000 cores that are used for graphical calculations, however the architecture can now be exposed for non-graphical applications with APIs such as CUDA~\cite{cudapg} and OpenCL~\cite{opencl08}. Such \emph{general purpose graphics processing units} (GPGPU) can greatly outperform a CPU if a problem can be parallelized accordingly.

Because each event in a Monte Carlo sample is independent, oscillation weight calculations can be performed in parallel. The library \texttt{Prob3++}~\cite{prob3pp} was ported to the GPU using the \emph{compute unified device architecture} (CUDA) API to enable fine-grained concurrent calculations. The results displayed in figure~\ref{fig:compare} show the execution times for varying numbers of calculations in series (CPU) and parallel (GPU). Also compared is the original code running multithreaded using OpenMP~\cite{openmpcite}.

\subsection{Method}
In the results presented, a series of C/C++ algorithms for calculating oscillation probabilities were ported to CUDA. Functions that execute on the device must be compiled separately by the \emph{nvcc} compiler provided by NVIDIA and linked into the host program using a compiler such as \emph{gcc}. 

Within the GPU code, an array of energy values were allocated and instantiated in host memory (the system's RAM) and then copied to the device memory (the graphics card's video RAM) using API function calls provided by CUDA.

In addition to the event energies, components that are dependent only on the oscillation parameters (i.e. Equation 10 of~\cite{bib1}) are computed on the CPU and then copied to the GPU in the same manner as the energy array.

The calculations in \texttt{Prob3++} were modified into a set of CUDA kernel functions (functions that run in parallel on the GPU) and were then executed on each element of the array in parallel, which performs the oscillation probability calculation in double precision. The result of this calculation is written to an array in the device memory, and is then copied back to the host. All memory allocation and transfer operations to and from the GPU device are handled via CUDA API functions. A simplified example of this process can be found in listing~\ref{lst:code}.

\begin{lstlisting}[label=lst:code,caption=Example of copying data to GPU memory and executing a kernel.]
// size of array
size_t size = n * sizeof(double);

// allocate host memory
double *true_energy_host = (double*) malloc(size);
double *osc_weight_host = (double*) malloc( size);

// allocate device memory
double *true_energy_dev = cudaMalloc((void **) &true_energy_device, size);
double *osc_weight_dev = cudaMalloc((void **) &osc_weight_device, size);

// fill energy array
...

// copy energy array to the device
cudaMemcpy(true_energy_dev, true_energy_host, size, cudaMemcpyHostToDevice);

// instantiate and perform copy of mixing matrix
...

// execute GPU kernel on the array
calculateOscProb<<<gridsize, blocksize>>>(...);

// copy the results back to the host
cudaMemcpy(osc_weight_host, osc_weight_dev , size, cudaMemcpyDeviceToHost);
\end{lstlisting}

\subsection{Results and Validation}
\begin{figure}[htp!]
\centering
\includegraphics[scale=0.65]{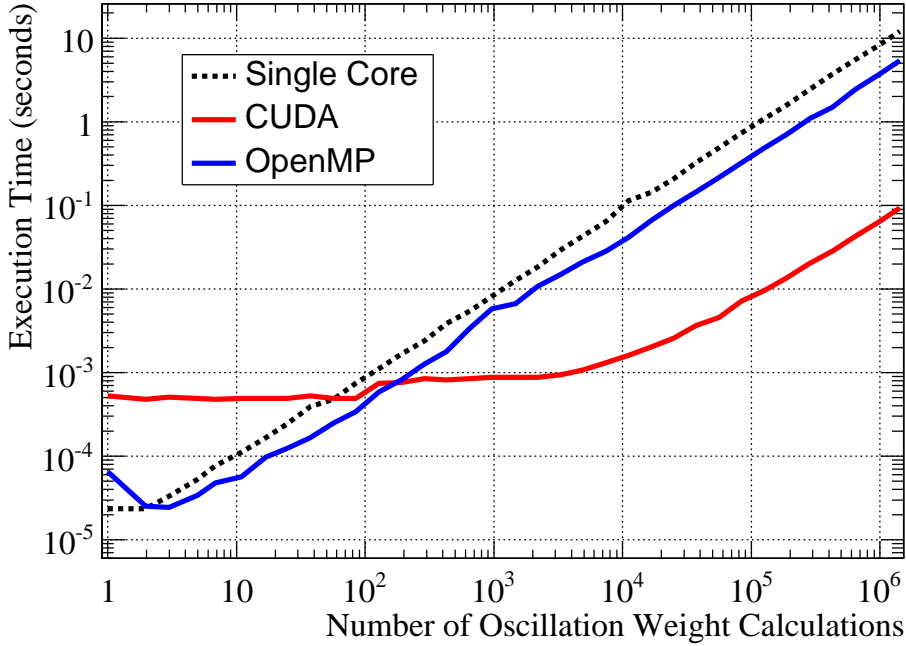}
\caption{Comparison of execution time for varying numbers of calculations between CPU and GPU implementations. The plateau observed in the CUDA results is due to the total number of threads not yet fully occupied. At $10^3$-$10^4$ number of calculations, the GPU becomes saturated and starts to execute in series.}
\label{fig:compare}
\end{figure}

\begin{figure}[hbp!]
\centering
\includegraphics[scale=0.65]{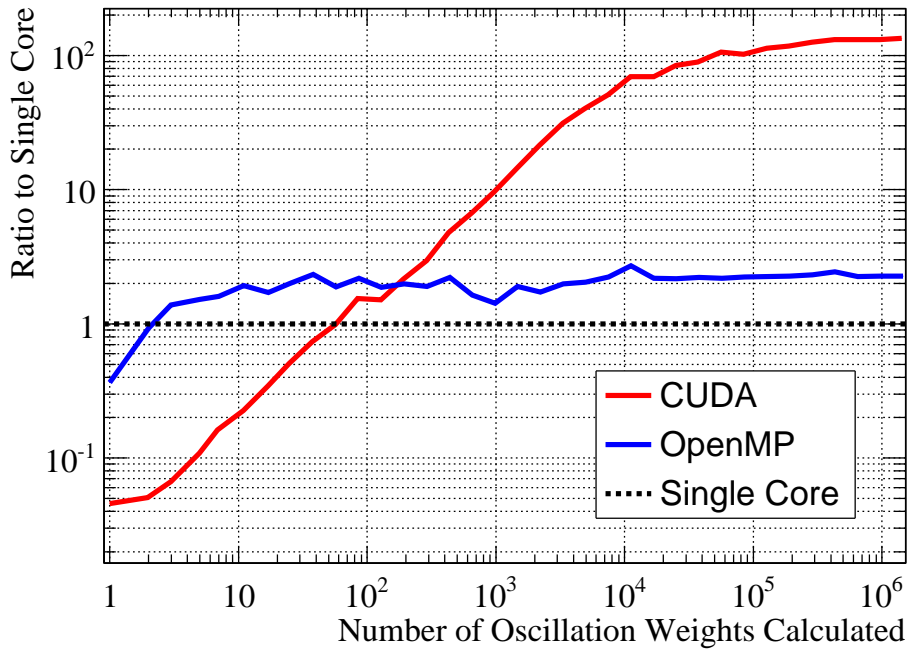}
\caption{Execution time plotted as a ratio to the single core implementation.}
\label{fig:ratio}
\end{figure}

The Comparison of CPU vs. GPU execution times as a function of number of events reweighted shows the CPU performing better at small number of events, with the GPU performing up to 132 times faster at 1.45 million calculations (figure~\ref{fig:compare}). The "crossover" point is hardware dependent, and is expected to change with different CPU/GPU combinations, and also different algorithm implementations. At best, the multi-threaded code gains only 2-3 times speed improvement. figure~\ref{fig:ratio} shows the benchmark with results plotted as a ratio to single core execution time. As seen in figures~\ref{fig:compare} and~\ref{fig:ratio}, the GPU implementation plateaus until it reaches a point where all threads are occupied and the limit of concurrent execution is reached~\cite{plateau}.

 The overheads associated with copying to and from host and device memory across the PCI-E bus can be a large source of latency, and as can be seen in figure~\ref{fig:compare}, the CPU will outperform the GPU if the number of concurrent calculations is small.


\begin{figure}
\centering
\includegraphics[scale=0.65]{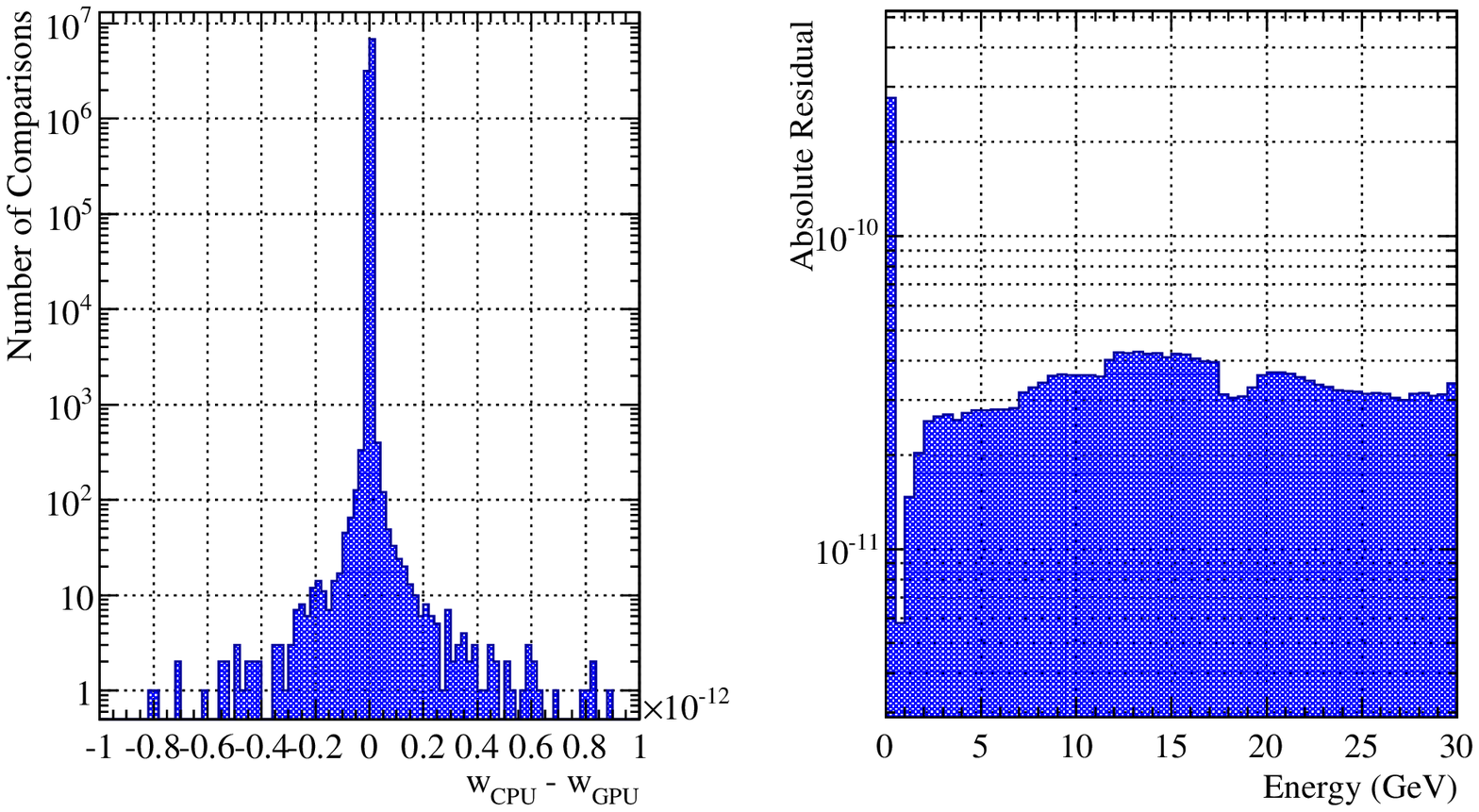}
\caption{\textbf{Left:} Residuals between weights calculated on CPU $w_{CPU}$ and GPU $w_{GPU}$ for the same oscillation parameters and value of energy. \textbf{Right:} The absolute difference between energy spectra weighted by $w_{CPU}$ and $w_{GPU}$.}
\label{fig:resid}
\end{figure}

To validate the GPU code, 10 million random energy values were drawn from a uniform distribution between 0 and 30 GeV, and were used to calculate oscillation weights on CPU and GPU. The residuals between CPU and GPU calcuations were found to be on the order of  $10^{-12}$ for double precision, and are plotted in figure~\ref{fig:resid}. The residual is attributed to the difference between hardware implementations of arithmetic operations~\cite{cuda_precision}, and in this test is considered negligible.

The GPU implementation and original version of \texttt{Prob3++} were also compared within a simple toy oscillation fitter written using the \emph{Bayesian analysis toolkit}~\cite{bat}. The motivation is to give realistic measure of speed improvement for an application in a physics analysis, as well as to show that there is negligible difference between both CPU and GPU methods when used in a realistic way. The fit uses a Markov Chain Monte Carlo to sample the oscillation parameter space, building a Bayesian posterior density via the Metropolis Hastings algorithm, from which credible intervals can be constructed.
The likelihood function is defined as:

\begin{equation}
L(\vec{o},\vec{f}| \vec{D}) = \prod\limits_{i}p(\vec{D}|\vec{o},\vec{f})
\end{equation}

Where $\vec{o}$ are the two parameters of interest $\theta_{23}$ and $\Delta m^2_{32}$, $\vec{f}$ are the nuisance parameters $\theta_{12}, \theta_{13}, \Delta m^2_{12}$ and $\delta_{cp}$, and $p$ is the probability mass function of a dataset $\vec{D}$ given parameters $\vec{o}$ and $\vec{f}$. The toy fit simulates a long baseline $\nu_{\mu}$ disappearance analysis by fitting a fake $\nu_{\mu}$ far-detector energy spectra $\vec{D}$, created by sampling from a landau function and weighted using the oscillation parameters found in Table~\ref{tab:osc_table}.

The PDF is constructed by taking a large number of samples (on the order of millions) from the landau distribution and binning these samples into a histogram weighted by the oscillation probability calculated with \texttt{Prob3++}. An example of oscillated and unoscillated spectra can be seen in figure~\ref{fig:oscprob}. 

As the Markov Chain Monte Carlo proposes a new set of oscillation parameters each step, the PDF is reconstructed using the event-by-event method described above and compared to the data. Therefore the calculation of oscillation weights provides a large overhead to the fit method and is directly related to the calculation of likelihood.

 The 5 oscillation parameters have flat prior distributions and thus have no likelihood constraint term, and all parameters are fixed at the values listed in Table~\ref{tab:osc_table} except $\theta_{23}$ and $\Delta m^2_{32}$ which are free to float.

\begin{figure}
\centering
\includegraphics[scale=0.65]{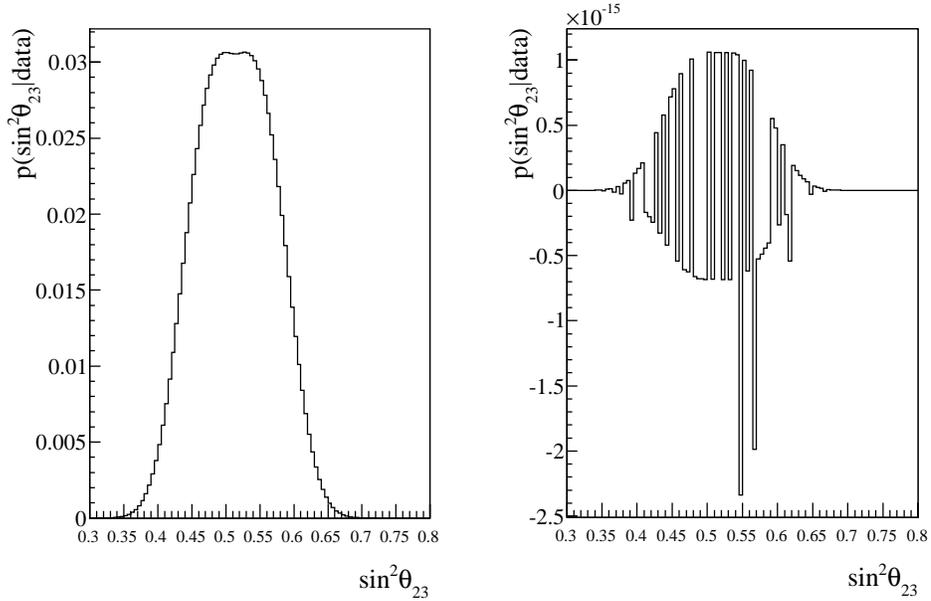} 
\caption{\textbf{Left:} 1-dimensional  $\sin^{2}(\theta_{23})$ marginal distribution. \textbf{Right:} Difference between the 1-dimensional marginal distribution of $\sin^{2}(\theta_{23})$ generated on CPU and GPU. The marginal distribution encodes information about the most probable value and the uncertainty of the parameter.}
\label{fig:marg_diff}
\end{figure}

The best fit and error value of the fitter was compared between CPU and GPU oscillation reweighting methods.
The difference between CPU and GPU made spectras and posterior distributions using identical oscillation parameters was found to be to an acceptable precision, and plotted in figure~\ref{fig:marg_diff}. Furthermore, an order of magnitude speed increase was observed for the overall fitting procedure by off-loading oscillation reweighting to the GPU. 

The results presented are prepared using an Intel Xeon E5640 quad-core processor running at 2.67 GHz, and an NVIDIA M2070 GPU with 448 CUDA cores running at 1.15 GHz. The code is compiled for 64-bit hardware using the gcc compiler version 4.6.3 with the -O2 optimization flag, and the CUDA toolkit version 5. OpenMP code is restricted to use 4 threads which ensures execution on the physical cores of the CPU.

\section{Conclusion}
The parallel implementation of oscillation reweighting enables the improvement of neutrino analyses via the computation of Monte Carlo weights on an event-by-event basis, which is a limiting factor of an analysis if performed soley on a CPU. Event-by-event reweighting retains all the Monte Carlo spectral shape information that is otherwise lost when binned into an histogram. More importantly, by being able to discriminate events within a sample of Monte Carlo, event migrations can be modelled, and as statistics of neutrino experiments increases this systematic effect will become more prominent. This has scope in current long-baseline neutrino experiments like T2K and NO$\nu$A, and future ones such as LBNE.

The CUDA implementation of \texttt{Prob3++} is available at the following web address: 

\begin{center}
\texttt{\href{http://hep.ph.liv.ac.uk/~rcalland/probGPU}{http://hep.ph.liv.ac.uk/~rcalland/probGPU}}
\end{center}

\acknowledgments

The author would like to thank R. Wendell for providing the original \texttt{Prob3++} library, the Liverpool High Energy Physics computing staff for their support, and the T2K experiment for access to official Monte Carlo and oscillation analysis software, from which this study was inspired.


\bibliography{ref}{}
\bibliographystyle{ieeetr}

\end{document}